\documentclass[twocolumn,prl,aps,showpacs,bibnotes]{revtex4}

\usepackage{amsmath}
\usepackage{verbatim}
\usepackage{graphicx}
\usepackage[usenames]{color}
\usepackage{psfrag}
\usepackage[ps2pdf,colorlinks,bookmarks]{hyperref}

\definecolor{linkblue}{rgb}{0,0,0.8}
\definecolor{linkgreen}{rgb}{0,0.5,0}
\hypersetup{pdfpagemode=None, pdfstartview=FitH, linkcolor=linkblue, %
            citecolor=linkgreen, urlcolor=linkblue}

\def\beq{\begin{equation}}
\def\eeq{\end{equation}}
\def\bea{\setlength\arraycolsep{1.4pt}\begin{eqnarray}}
\def\eea{\end{eqnarray}}
\def\bit{\begin{itemize}}
\def\eit{\end{itemize}}

\def\ie{{i.e.}}
\def\eg{{e.g.}}

\def\fig{Fig.~}

\def\pd{\partial}

\def\ld{\left}
\def\rd{\right}

\def\fr{\frac}

\def\camb{\textsc{camb}}
\def\cosmomc{\textsc{cosmomc}}
\def\Heds{H^{\rm EdS}}

\def\Teds{T_0^{\rm EdS}}
\def\zeds{z^{\rm EdS}}

\def\lbao{l^{\rm BAO}}
\def\Dtbao{\Delta\theta^{\rm BAO}}
\def\Dzbao{\Delta z^{\rm BAO}}

\begin{document}

\title{Can we a{\em void} dark energy?}

\author{James P. Zibin} \email{zibin@phas.ubc.ca}
\author{Adam Moss} \email{adammoss@phas.ubc.ca}
\author{Douglas Scott} \email{dscott@phas.ubc.ca}
\affiliation{Department of Physics \& Astronomy, %
University of British Columbia, %
Vancouver, BC, V6T 1Z1  Canada}

\date{\today}

\begin{abstract}
The idea that we live near the centre of a large, nonlinear void 
has attracted attention recently as an alternative to dark energy or 
modified gravity.  We show that an appropriate void profile can 
fit both the latest cosmic microwave background and supernova data.  
However, this requires either a fine-tuned primordial spectrum or a 
Hubble rate so low as to rule these models out.  We also show that 
measurements of the {\em radial\/} baryon acoustic scale can 
provide very strong constraints.  Our results present a 
serious challenge to void models of acceleration.
\end{abstract}


\maketitle

{\em Introduction.---}%
The last decade has seen the solidification of the 
Standard Model of Cosmology (SMC; see, \eg, \cite{scott06}), which contains 
about $75\%$ dark energy driving the acceleration of a 
flat, homogeneous and isotropic Friedmann-Lema\^itre-Robertson-Walker 
(FLRW) background.  A very broad range of observations are consistent 
with the SMC, including cosmic microwave background (CMB) (see, \eg, 
\cite{Hinshaw:2008kr}), Type Ia supernovae 
(SNe Ia) (\eg\ \cite{union08}), baryon acoustic oscillations (BAO) 
(\eg\ \cite{pcenpps07}), integrated Sachs-Wolfe (ISW) effect correlations 
(\eg\ \cite{hhpsb08}), weak lensing studies (\eg\ \cite{fuetal08}), etc.

   Given this impressive convergence of observations, it would be very 
surprising if the data still fit an alternative model, which lacked two of 
the main planks of the SMC.  Yet just such a radical proposal has attracted 
considerable attention recently.  The idea is to drop the dark 
energy and the Copernican principle, and instead suppose 
that we are near the centre of a large, nonlinearly underdense,
nearly spherical {\em void\/} surrounded by a flat, matter dominated 
Einstein-de~Sitter (EdS) spacetime (\cite{celerier99}; see \cite{enqvist08} 
for a brief review).  
By tuning the radial void profile, it is possible to match the luminosity 
distance-redshift relation of the SMC \cite{cr06ykn08}, so if these 
models are to be ruled out, we need more than just the SN data.  
Recently, constraints from the CMB acoustic scale, BAO scale, and Hubble 
rate have been placed on void models, although, remarkably, they 
have not yet been ruled out \cite{abnv07,gbh08bw08}.

   We must stress that void models contain two elements which 
appear extremely unlikely within the standard cosmological framework, 
and for which no viable explanations have been proposed.  Voids of the 
size required to fit the SN data (hundreds of Mpc to Gpc scales) 
correspond to fluctuations 
of very many $\sigma$ in standard structure formation scenarios \cite{hs08}.  
Also, we must be very close to the void centre to avoid a large CMB 
dipole \cite{aa06}.

   However, as overwhelming as these difficulties may appear, they 
are essentially philosophical in nature, and so it would be 
important to rule these models out on the basis of observations.  
With this goal in mind we confront void models with 
several sets of current data, providing three main advances over 
previous studies.  Firstly, we allow for a wide range of void 
profiles, employing a spline parameterization.  Secondly, we carefully 
calculate the CMB anisotropy spectrum that a void observer would see 
and confront the {\em full spectrum\/} with recent data sets.  
Finally, we show that the {\em radial\/} BAO scale is a powerful 
discriminator between void and standard models.

{\em Specifying the void model.---}%
We model the contents of the Universe 
at late times as pressureless matter, with density $\rho$.  Observations 
are consistent with isotropy, so we place 
the observer at the centre of spherical symmetry.  The exact solution to 
Einstein's equations in this case is known as the Lema\^itre-Tolman-Bondi 
(LTB) spacetime \cite{lemaitre33tolman34bondi47}.  It is described by 
two free radial functions, which correspond 
to the growing and decaying modes in the limit of small perturbations 
about FLRW \cite{silk77,z08}.  As stressed in \cite{z08}, it is crucial 
to consider only voids with vanishing decaying mode (\ie\ uniform 
``bang time''), if we are to be able to specify the initial 
conditions (ICs) for perturbations at early times.  This is because 
in this case the void {\em itself\/} will be a small 
perturbation from FLRW at early times, and standard inflationary ICs can 
be specified on top of the void.  Since our analysis will 
include the BAO perturbation scale (which is set before recombination) 
evaluated {\em inside\/} the void, we must assume vanishing decaying mode 
in our work.  Thus only one radial function is required to specify 
the void model.

   The LTB spacetime can be described by the metric
\beq
ds^2 = -dt^2 + \fr{Y'^2}{1 - K}dr^2 + Y^2d\Omega^2.
\eeq
Here $Y(t,r)$ and $K(r)$ are determined by the exact LTB solution (see, 
\eg, \cite{z08}) once the single free radial function is specified, and 
$Y' \equiv \pd Y/\pd r$, where $r$ is a comoving radial coordinate.  
Errors due to ignoring radiation are inevitable with the LTB 
solution, but we estimate our results are accurate to the percent level 
or better \cite{voidsinprep}.

   We chose to define the radial profile in terms of the comoving 
density perturbation, $\delta\rho(t_i,r)/\rho(t_i)$, specified at the 
early time $t_i$.  
To be confident that no important regions 
of ``profile space'' are missed, we fit a three-point cubic spline to the 
initial density fluctuation $\delta\rho_j \equiv \delta\rho(t_i,r_j)$, 
where  $j = 1$, $2$, $3$. We fix $r_1 = 0$ and enforce the void to be 
smooth at the origin and to smoothly match to EdS at large $r$ 
by setting $\delta\rho'(r_1) = \delta\rho'(r_3) = \delta\rho_3 = 0$.  
This leaves a total of four free parameters: the density at 
the origin, $\delta \rho_1$, the density and radius at the midpoint, 
$\delta \rho_2, r_2$, and the radius $r_3$ at which we match to EdS.  
We have checked that additional spline points do not significantly improve 
the fit to current data.  We consider two classes of profiles: 
``constrained,'' for which we impose $\int\delta\rho(t_i) r^2 dr \leq 0$, 
and ``unconstrained,'' which are free.

   A void model is completely specified by the radial profile, 
the Hubble rate at the void centre today, $H_0$, the 
radiation density today, which is fixed by the CMB mean temperature, 
$T_0 = 2.725$~K \cite{mfsmw99}, and the baryon fraction $f_b \equiv 
\rho_b/\rho_m$.  Outside the void we asymptote to EdS.

\begin{figure}[ht]\begin{center}
\includegraphics[width=\columnwidth]{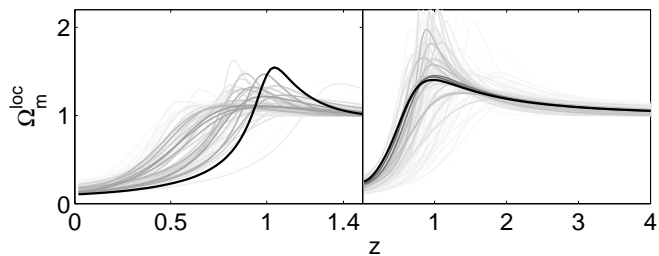}
\caption{Local density parameter $\Omega^{\rm loc}_m$ versus redshift $z$ for 
constrained 
(left) and unconstrained (right) voids.  The profiles are sampled from 
MCMC chains and the grayscale level indicates the relative likelihood.}
\label{profilefig}
\end{center}\end{figure}

   In \fig\ref{profilefig}, we plot several void profiles on the 
observer's past light cone, in terms of the local density parameter 
$\Omega^{\rm loc}_m \equiv 24\pi G\rho/\theta^2$, where $\theta$ is the 
comoving 
expansion \cite{z08} (this definition reduces to the standard density 
parameter in the FLRW case).  The profiles are sampled from 
a Markov-Chain-Monte-Carlo (MCMC) process, fitting to SN+CMB data as 
described below.  In these plots, values $\Omega^{\rm loc}_m < 1$ correspond to the 
central void, while $\Omega^{\rm loc}_m > 1$ indicates an overdense shell region.  
The constrained voids tend to be essentially ``compensated,'' while the 
unconstrained tend to be strongly ``overcompensated.''

{\em Cosmic microwave background.---}%
The CMB anisotropy power spectrum, $C_\ell$, contains much 
information, and so can potentially 
provide strong constraints on void models.  Two main 
factors determine the detailed {\em shape\/} of the $C_\ell$ spectrum: 
the {\em primordial\/} perturbation spectrum, which, in the 
simplest models of inflation, is close to scale 
invariant and essentially featureless; and the local 
physics during recombination, which is determined by the 
composition of the Universe at that time.  The {\em angular scale\/} at 
which features of fixed physical scale appear 
is determined by the physical circumference of the observer's last 
scattering surface (LSS), which is affected by the geometry of 
the Universe and the time of observation \cite{zms07}.

   Once the void model has been specified, it is straightforward to 
calculate the $C_\ell$ spectrum at all but the largest scales.  The 
anisotropies generated at the LSS cannot be affected by the void for 
an on-centre observer, because of the isotropy of the LTB 
background \cite{note1}.  
Our procedure to calculate the $C_\ell$'s is simply to find the 
parameters for an effective EdS model with the same physics at 
recombination and LSS physical circumference 
as the void model.  Then those effective parameters can be fed 
into public CMB anisotropy codes---we used \camb\ \cite{lcl00notecamb}.

   Explicitly, we numerically integrate along the past light cone from 
the void centre today ($r = z = 0$) to a redshift $z_m$ to find the 
corresponding coordinates $(t_m,r_m) = (t(z_m),r(z_m))$ using the exact 
LTB relations
\beq
\fr{dr}{dz} = \fr{\sqrt{1 - K}}{(1 + z){\dot Y}'}, \qquad
\fr{dt}{dz} = \fr{-Y'}{(1 + z){\dot Y}'}.
\eeq
We choose $z_m$ large enough so that $r_m$ is sufficiently far outside 
the void that spatial curvature (and shear) is negligible there, but not 
so large that radiation is important at $t_m$ at background level.  In 
practice, values $z_m \simeq 100$ meet these criteria.  We 
also evaluate the Hubble rate, $H_m$, at $z_m$ using the exact LTB 
solution.  Finally, we set the spatial curvature to precisely zero, and 
integrate back up the light cone {\em into EdS\/} to comoving coordinate 
$r^{\rm EdS} = 0$, using the FLRW relation $d\zeds/dr^{\rm EdS} = \Heds(z)$.  
The correct LSS circumference is ensured by using the relation 
$Y(t_m,r_m) = a(r^{\rm EdS}_m)r^{\rm EdS}_m$, for FLRW scale factor $a$, 
to set the IC for this integration.  The result of integration, $\zeds_m$, 
allows us to calculate the effective EdS mean temperature and Hubble 
rate via
\beq
\!\!\Teds = T_0(1 + \zeds_m),\quad
\Heds_0 = H_m\ld(\fr{1 + \zeds_m}{1 + z_m}\rd)^{3/2}.
\eeq
The parameters 
$\Teds$, $\Heds_0$, $\Omega_m = 1$, $\Omega_\Lambda = \Omega_K = 0$, and 
$f_b$ are then fed into \camb\ to calculate the $C_\ell$ spectrum which 
would be observed in the specified void.  Note that the effective 
parameters $\Teds$ and $\Heds_0$ will generally differ from the true 
temperature and Hubble rate observed at the void centre today, 
$T_0 = 2.725$~K and $H_0$.

   Importantly, we find that the effective set $\Teds = 3.372$~K, 
$\Heds_0 = 51.0$~km$\,$s$^{-1}\,$Mpc$^{-1}$, and $f_b = 0.165$ produce 
a $C_\ell$ spectrum that matches that of the Wilkinson Microwave 
Anisotropy Probe (WMAP) best-fit $\Lambda$ model \cite{Hinshaw:2008kr} 
at all but the largest scales.  A 
void model must have effective parameters close to these if it is to fit 
the CMB.

{\em Data fitting and results.---}%
We used \cosmomc\ \cite{lb02} to create MCMC chains to 
estimate confidence limits on the parameters.  Along with the four void 
spline quantities, the basic cosmological parameters used in the fit are 
$f_b$, $H_0$, and the amplitude $A_s$ and spectral index $n_s$ of 
adiabatic initial perturbations \cite{note1}.

   SNe Ia provide important evidence for acceleration within the 
standard FLRW framework, and hence we must ensure that our voids fit these 
observations.  We used the recent Union compilation \cite{union08}, 
consisting of $307$ SNe with $z = 0.015$--$1.55$ \cite{noteunion}.  We also 
used CMB data, namely the $5$-year WMAP 
results \cite{Hinshaw:2008kr} and four 
experiments at higher resolution: 
ACBAR \cite{Reichardt:2008ay}, 
Boomerang \cite{Jones:2005yb}, CBI \cite{Readhead:2004gy}, and 
QUaD \cite{Pryke:2008xp}.  We also applied a conservative prior that 
$\Omega^{\rm loc}_m > 0.1$ at the void centre.

\begin{figure}
\begin{center}
\includegraphics[width=\columnwidth]{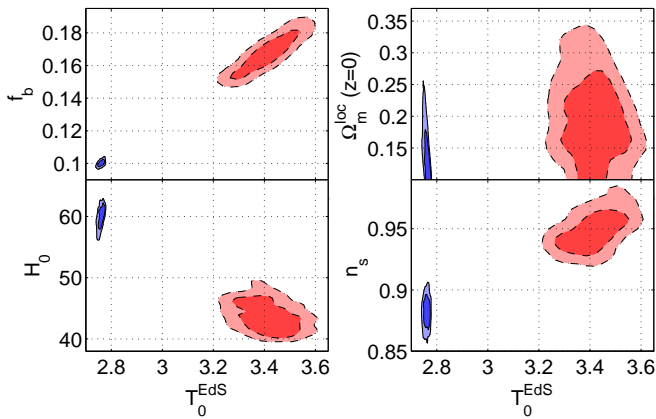}
\caption{Likelihood contours at $68\%$ and $95\%$ confidence for 
constrained (blue, solid contours) 
and unconstrained (red, dashed contours) voids.}
\label{likelyfig}
\end{center}
\end{figure}

   In \fig\ref{likelyfig} we show a selection of 2D likelihood surfaces 
for the constrained and unconstrained void models.  The parameters 
are very different in each case.  For constrained voids, the 
effective temperature, $\Teds = 2.760 \pm 0.008$~K, is similar to $T_0$.  
This $\Teds$ is far too low to provide a good fit 
to the observed CMB spectrum---we find $\Delta \chi^2=162$ between the 
constrained void and $\Lambda$ model for CMB+SN data, 
with almost all of this difference coming from the 
CMB.  This poor fit also leads to very low $f_b = 0.100 \pm 0.001$ and 
$n_s = 0.88 \pm 0.01$.

   For the unconstrained voids, however, much higher effective temperatures 
are possible, due to the geometrical effect of the overdense shell.  
These temperatures are sufficiently high to fit the CMB 
well---the fit to CMB+SN is actually slightly {\em better\/} than $\Lambda$, 
with $\Delta \chi^2=-1.4$.  However, this requires an unusually low local 
Hubble rate of $H_0 = 44 \pm 2$ $\rm km \, s^{-1} \, Mpc^{-1}$, as 
\fig\ref{likelyfig} shows.  Recent {\em local} estimates range between 
$57$ and $79$ $\rm km \, s^{-1} \, Mpc^{-1}$ at $1 \sigma$ 
\cite{Jackson:2007ug}, and so this class of void model is ruled out at 
high confidence.

\begin{figure}[ht]\begin{center}
\includegraphics[width=\columnwidth]{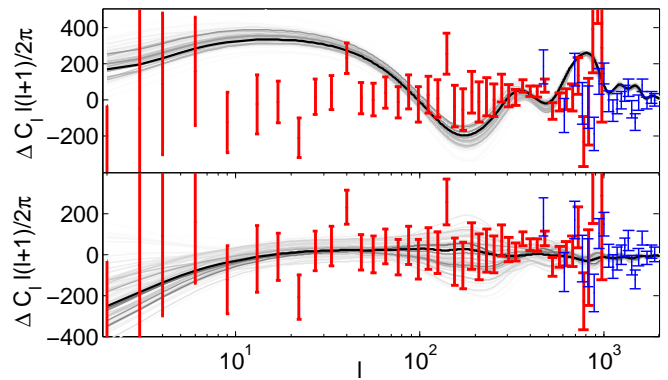}
\caption{Residual $C_\ell$'s (in $\mu$K$^2$) from the best-fit $\Lambda$ 
model for 
constrained (top) and unconstrained (bottom) voids.  Binned WMAP 
data are shown by the red (thick) error bars, and ACBAR data by the blue 
(thin) bars.}
\label{cmbfig}
\end{center}\end{figure}

   In \fig\ref{cmbfig} we present the residuals of the void $C_{\ell}$'s 
from the best-fit $\Lambda$ model.  The best-fit void model is shown by 
the black curve, along with $100$ other spectra sampled randomly from 
the MCMC chains.  The grayscale level 
indicates the relative likelihoods.  It is clear that in the constrained 
case one cannot fit the CMB data without introducing fine-tuned 
features to the primordial spectrum, since the physics at recombination 
is wrong.
 
{\em Baryon acoustic scale.---}%
The physics before recombination imprints a fixed comoving scale 
into the matter power spectrum in the form of the BAO scale, carrying 
much useful information about the geometry and expansion 
history assuming an FLRW background (see, \eg, \cite{se03}).  
It is therefore important to assess the usefulness of BAO data in constraining 
void models for acceleration.

   The first step is to 
evaluate the {\em physical\/} BAO scale, $\lbao_i$ (also called the 
sound horizon at the drag epoch), at some time $t_i$ 
early enough that the void background is well approximated by FLRW.  
To do this, we find an effective EdS model with the same physics at 
early times as the specified void model using the same procedure 
used to calculate the void $C_\ell$'s, except that it is not necessary 
here to match the effective and true LSS circumferences.  Then we 
calculate $\lbao_i$ using the fitting function from \cite{eh98} applied 
to the effective EdS parameters.  Next, $\lbao_i$ is propagated up to 
redshift $z$ on the void observer's past light cone using the 
LTB metric.  The background 
shear causes the {\em physical\/} BAO scales in the transverse and 
radial directions to differ; they are given, respectively, by
\beq
\lbao_\perp(z) = \fr{\lbao_iY(z)}{Y(t_i,r(z))},\quad
\lbao_\parallel(z) = \fr{\lbao_iY'(z)}{Y'(t_i,r(z))}.
\eeq
Here $Y(z) \equiv Y(t(z),r(z))$, and similarly for $Y'(z)$.  
BAO observations are often expressed as a BAO length scale today, but 
such values necessarily depend on the assumed background.  Instead, a 
model-independent expression of the transverse and radial BAO (RBAO) scales 
is in terms of the corresponding angular and redshift increments,
\beq
\!\!\!\Dtbao(z) = \fr{\lbao_\perp(z)}{Y(z)},\,\,\,\,
\fr{\Dzbao(z)}{1 + z} = \lbao_\parallel(z)\fr{\dot{Y'}(z)}{Y'(z)}.
\eeq
Reference \cite{cbl08} emphasized the importance of distinguishing 
radial and angular scales in this context.

\begin{figure}[ht]\begin{center}
\includegraphics[width=\columnwidth]{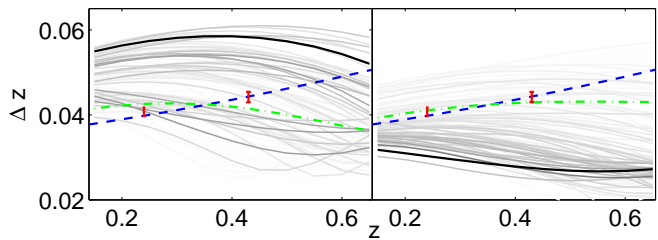}
\caption{RBAO scale $\Dzbao$ for constrained (left) and unconstrained 
(right) voids.  Also shown are the RBAO data from \cite{gch08gms08} (error 
bars), best-fit $\Lambda$ model (blue, dashed), and best-fit void 
including RBAO constraints (green, dot-dashed).}
\label{baofig}
\end{center}\end{figure}

   Importantly, $\Dzbao$ contains two factors that {\em reinforce\/} 
each other in the peripheral void region.  The quantity 
$\dot{Y'}(z)/Y'(z)$ is the expansion rate in the radial direction, which 
is suppressed in the overdense periphery.  This, in turn, results in 
a suppressed RBAO scale, $\lbao_\parallel(z)$, in the 
periphery.  The net effect is a strong suppression of $\Dzbao$ in this 
region compared with the standard $\Lambda$ case, as illustrated in 
\fig\ref{baofig} for a set of profiles from the same MCMC chain as we 
obtained above \cite{note2}.

   Recently, RBAO measurements have been presented \cite{gch08gms08} for the 
observable $\Dzbao$ rather than the model-dependent $\lbao$.  
Figure~\ref{baofig} shows that the new data are in excellent agreement with 
standard $\Lambda$, but strongly at odds with the likeliest void models.  
Including the new RBAO data 
points worsens our best fits by $\Delta\chi^2 = 16$ 
(constrained profiles) and $9.8$ (unconstrained).  Such large 
$\Delta\chi^2$ values for only two extra data points show that 
the new RBAO technique is {\em already\/} a considerable obstacle to void 
models.

{\em Discussion.---}%
We have concentrated here on constraints from SN, CMB, and RBAO data.  
In future work \cite{voidsinprep} we will also consider the independent 
constraints from the {\em amplitude\/} of matter fluctuations \cite{z08}.  
These techniques all rely on observations confined to our past 
light cone.  Very promising are measurements that sample the {\em 
interior\/} of the light cone, via spectral distortions 
\cite{goodman95cs07} or the kinematic Sunyaev-Zel'dovich effect 
\cite{gbh08b} (see also \cite{uce08} for a distinct approach).

   Nevertheless, we have already shown here that an appropriate 
overcompensated void profile {\em can\/} fit both the SN and CMB data, 
but at the expense of an $H_0$ so low that it can be ruled out.  
Constrained voids provide a better $H_0$, but a very poor fit to the 
CMB without fine tuning of the primordial spectrum.  We stress the 
significance of this result:  it is an extraordinary achievement of the 
SMC that it predicts the detailed shape of the $C_\ell$ 
spectrum using only a few parameters.  Losing this predictive power and 
requiring a fine-tuned primordial spectrum is a severe price to pay 
for the allure of $\Lambda = 0$.

   We also showed that 
early RBAO results already impose very strong constraints on void 
models.  In particular, RBAO poses a serious challenge to the constrained 
profiles which is free of any subjectivity that some may argue is 
inherent in the above fine-tuning argument.  Future BAO surveys such as PAU 
\cite{pau08} and BOSS \cite{noteboss} will provide improved precision 
out to greater redshifts, placing unprecedented constraints on 
inhomogeneity.  In this era of precision cosmology we can in 
fact begin to test the Copernican principle.

{\em Acknowledgments.---}%
This research was supported by the Natural Sciences and Engineering Research 
Council of Canada, the Canadian Foundation for Innovation, and the British 
Columbia Knowledge Development Fund.  We thank 
Kris Sigurdson, Troels 
Haugb\o lle, Alessio Notari, and Tirthabir Biswas for useful discussions.

\bibliography{bib}

\begin{thebibliography}{45}
\expandafter\ifx\csname natexlab\endcsname\relax\def\natexlab#1{#1}\fi
\expandafter\ifx\csname bibnamefont\endcsname\relax
  \def\bibnamefont#1{#1}\fi
\expandafter\ifx\csname bibfnamefont\endcsname\relax
  \def\bibfnamefont#1{#1}\fi
\expandafter\ifx\csname citenamefont\endcsname\relax
  \def\citenamefont#1{#1}\fi
\expandafter\ifx\csname url\endcsname\relax
  \def\url#1{\texttt{#1}}\fi
\expandafter\ifx\csname urlprefix\endcsname\relax\def\urlprefix{URL }\fi
\providecommand{\bibinfo}[2]{#2}
\renewcommand{\eprint}[1]{arXiv:\href{http://arxiv.org/abs/#1}{#1}}
\providecommand{\doi}[2]{\href{http://dx.doi.org/#1}{#2}}

\bibitem[{\citenamefont{{Scott}}(2006)}]{scott06}
\bibinfo{author}{\bibfnamefont{D.}~\bibnamefont{{Scott}}},
  \bibinfo{journal}{Can. J. Phys.}
  \textbf{\doi{10.1139/P06-066}{\bibinfo{volume}{84}}},
  \doi{10.1139/P06-066}{\bibinfo{pages}{419}} (\bibinfo{year}{2006}),
  \eprint{astro-ph/0510731}.

\bibitem[{\citenamefont{Hinshaw et~al.}(2008)}]{Hinshaw:2008kr}
\bibinfo{author}{\bibfnamefont{G.}~\bibnamefont{Hinshaw}} \bibnamefont{et~al.}
  (\bibinfo{collaboration}{WMAP}) (\bibinfo{year}{2008}), \eprint{0803.0732}
  [astro-ph].

\bibitem[{\citenamefont{Kowalski et~al.}(2008)}]{union08}
\bibinfo{author}{\bibfnamefont{M.}~\bibnamefont{Kowalski}} \bibnamefont{et~al.}
  (\bibinfo{year}{2008}), \eprint{0804.4142} [astro-ph].

\bibitem[{\citenamefont{Percival et~al.}(2007)}]{pcenpps07}
\bibinfo{author}{\bibfnamefont{W.~J.} \bibnamefont{Percival}}
  \bibnamefont{et~al.}, \bibinfo{journal}{Mon. Not. Roy. Astron. Soc.}
  \textbf{\doi{10.1111/j.1365-2966.2007.12268.x}{\bibinfo{volume}{381}}},
  \doi{10.1111/j.1365-2966.2007.12268.x}{\bibinfo{pages}{1053}}
  (\bibinfo{year}{2007}), \eprint{0705.3323} [astro-ph].

\bibitem[{\citenamefont{Ho et~al.}(2008)}]{hhpsb08}
\bibinfo{author}{\bibfnamefont{S.}~\bibnamefont{Ho}} \bibnamefont{et~al.},
  \bibinfo{journal}{Phys. Rev.}
  \textbf{\doi{10.1103/PhysRevD.78.043519}{\bibinfo{volume}{D78}}},
  \doi{10.1103/PhysRevD.78.043519}{\bibinfo{pages}{043519}}
  (\bibinfo{year}{2008}), \eprint{0801.0642} [astro-ph].

\bibitem[{\citenamefont{{Fu} et~al.}(2008)}]{fuetal08}
\bibinfo{author}{\bibfnamefont{L.}~\bibnamefont{{Fu}}} \bibnamefont{et~al.},
  \bibinfo{journal}{Astron. Astrophys.}
  \textbf{\doi{10.1051/0004-6361:20078522}{\bibinfo{volume}{479}}},
  \doi{10.1051/0004-6361:20078522}{\bibinfo{pages}{9}} (\bibinfo{year}{2008}),
  \eprint{0712.0884} [astro-ph].

\bibitem[{\citenamefont{Celerier}(2000)}]{celerier99}
\bibinfo{author}{\bibfnamefont{M.-N.} \bibnamefont{Celerier}},
  \bibinfo{journal}{Astron. Astrophys.} \textbf{\bibinfo{volume}{353}},
  \bibinfo{pages}{63} (\bibinfo{year}{2000}), \eprint{astro-ph/9907206}.

\bibitem[{\citenamefont{Enqvist}(2008)}]{enqvist08}
\bibinfo{author}{\bibfnamefont{K.}~\bibnamefont{Enqvist}},
  \bibinfo{journal}{Gen. Rel. Grav.}
  \textbf{\doi{10.1007/s10714-007-0553-9}{\bibinfo{volume}{40}}},
  \doi{10.1007/s10714-007-0553-9}{\bibinfo{pages}{451}} (\bibinfo{year}{2008}),
  \eprint{0709.2044} [astro-ph].

\bibitem[{\citenamefont{Chung and Romano}(2006)}]{cr06ykn08}
\bibinfo{author}{\bibfnamefont{D.~J.~H.} \bibnamefont{Chung}} \bibnamefont{and}
  \bibinfo{author}{\bibfnamefont{A.~E.} \bibnamefont{Romano}},
  \bibinfo{journal}{Phys. Rev.}
  \textbf{\doi{10.1103/PhysRevD.74.103507}{\bibinfo{volume}{D74}}},
  \doi{10.1103/PhysRevD.74.103507}{\bibinfo{pages}{103507}}
  (\bibinfo{year}{2006}), \eprint{astro-ph/0608403};
%
\bibinfo{author}{\bibfnamefont{C.-M.} \bibnamefont{Yoo}},
  \bibinfo{author}{\bibfnamefont{T.}~\bibnamefont{Kai}}, \bibnamefont{and}
  \bibinfo{author}{\bibfnamefont{K.-i.} \bibnamefont{Nakao}}
  (\bibinfo{year}{2008}), \eprint{0807.0932} [astro-ph].

\bibitem[{\citenamefont{Alexander et~al.}(2007)\citenamefont{Alexander, Biswas,
  Notari, and Vaid}}]{abnv07}
\bibinfo{author}{\bibfnamefont{S.}~\bibnamefont{Alexander}},
  \bibinfo{author}{\bibfnamefont{T.}~\bibnamefont{Biswas}},
  \bibinfo{author}{\bibfnamefont{A.}~\bibnamefont{Notari}}, \bibnamefont{and}
  \bibinfo{author}{\bibfnamefont{D.}~\bibnamefont{Vaid}}
  (\bibinfo{year}{2007}), \eprint{0712.0370} [astro-ph].

\bibitem[{\citenamefont{Garcia-Bellido and
  Haugboelle}(2008{\natexlab{a}})}]{gbh08bw08}
\bibinfo{author}{\bibfnamefont{J.}~\bibnamefont{Garcia-Bellido}}
  \bibnamefont{and}
  \bibinfo{author}{\bibfnamefont{T.}~\bibnamefont{Haugboelle}},
  \bibinfo{journal}{JCAP}
  \textbf{\doi{10.1088/1475-7516/2008/04/003}{\bibinfo{volume}{0804}}},
  \doi{10.1088/1475-7516/2008/04/003}{\bibinfo{pages}{003}}
  (\bibinfo{year}{2008}{\natexlab{a}}), \eprint{0802.1523} [astro-ph];
%
\bibinfo{author}{\bibfnamefont{K.}~\bibnamefont{Bolejko}} \bibnamefont{and}
  \bibinfo{author}{\bibfnamefont{J.~S.~B.} \bibnamefont{Wyithe}}
  (\bibinfo{year}{2008}), \eprint{0807.2891} [astro-ph].

\bibitem[{\citenamefont{Hunt and Sarkar}(2008)}]{hs08}
\bibinfo{author}{\bibfnamefont{P.}~\bibnamefont{Hunt}} \bibnamefont{and}
  \bibinfo{author}{\bibfnamefont{S.}~\bibnamefont{Sarkar}}
  (\bibinfo{year}{2008}), \eprint{0807.4508} [astro-ph].

\bibitem[{\citenamefont{Alnes and Amarzguioui}(2006)}]{aa06}
\bibinfo{author}{\bibfnamefont{H.}~\bibnamefont{Alnes}} \bibnamefont{and}
  \bibinfo{author}{\bibfnamefont{M.}~\bibnamefont{Amarzguioui}},
  \bibinfo{journal}{Phys. Rev.}
  \textbf{\doi{10.1103/PhysRevD.74.103520}{\bibinfo{volume}{D74}}},
  \doi{10.1103/PhysRevD.74.103520}{\bibinfo{pages}{103520}}
  (\bibinfo{year}{2006}), \eprint{astro-ph/0607334}.

\bibitem[{\citenamefont{{Lema{\^i}tre}}(1933)}]{lemaitre33tolman34bondi47}
\bibinfo{author}{\bibfnamefont{G.}~\bibnamefont{{Lema{\^i}tre}}},
  \bibinfo{journal}{Ann. Soc. Sci. Bruxelles} \textbf{\bibinfo{volume}{53}},
  \bibinfo{pages}{51} (\bibinfo{year}{1933});
%
\bibinfo{author}{\bibfnamefont{R.~C.} \bibnamefont{Tolman}},
  \bibinfo{journal}{Proc. Nat. Acad. Sci.} \textbf{\bibinfo{volume}{20}},
  \bibinfo{pages}{169} (\bibinfo{year}{1934});
%
\bibinfo{author}{\bibfnamefont{H.}~\bibnamefont{Bondi}}, \bibinfo{journal}{Mon.
  Not. Roy. Astron. Soc.} \textbf{\bibinfo{volume}{107}}, \bibinfo{pages}{410}
  (\bibinfo{year}{1947}).

\bibitem[{\citenamefont{{Silk}}(1977)}]{silk77}
\bibinfo{author}{\bibfnamefont{J.}~\bibnamefont{{Silk}}},
  \bibinfo{journal}{Astron. Astrophys.} \textbf{\bibinfo{volume}{59}},
  \bibinfo{pages}{53} (\bibinfo{year}{1977}).

\bibitem[{\citenamefont{Zibin}(2008)}]{z08}
\bibinfo{author}{\bibfnamefont{J.~P.} \bibnamefont{Zibin}},
  \bibinfo{journal}{Phys. Rev.}
  \textbf{\doi{10.1103/PhysRevD.78.043504}{\bibinfo{volume}{D78}}},
  \doi{10.1103/PhysRevD.78.043504}{\bibinfo{pages}{043504}}
  (\bibinfo{year}{2008}), \eprint{0804.1787} [astro-ph].

\bibitem[{\citenamefont{Moss et~al.}(2008)\citenamefont{Moss, Zibin, and
  Scott}}]{voidsinprep}
\bibinfo{author}{\bibfnamefont{A.}~\bibnamefont{Moss}},
  \bibinfo{author}{\bibfnamefont{J.~P.} \bibnamefont{Zibin}}, \bibnamefont{and}
  \bibinfo{author}{\bibfnamefont{D.}~\bibnamefont{Scott}}
  (\bibinfo{year}{2008}), \bibinfo{note}{in preparation}.

\bibitem[{\citenamefont{Mather et~al.}(1999)}]{mfsmw99}
\bibinfo{author}{\bibfnamefont{J.~C.} \bibnamefont{Mather}}
  \bibnamefont{et~al.}, \bibinfo{journal}{Astrophys. J.}
  \textbf{\doi{10.1086/306805}{\bibinfo{volume}{512}}},
  \doi{10.1086/306805}{\bibinfo{pages}{511}} (\bibinfo{year}{1999}),
  \eprint{astro-ph/9810373}.

\bibitem[{\citenamefont{Zibin et~al.}(2007)\citenamefont{Zibin, Moss, and
  Scott}}]{zms07}
\bibinfo{author}{\bibfnamefont{J.~P.} \bibnamefont{Zibin}},
  \bibinfo{author}{\bibfnamefont{A.}~\bibnamefont{Moss}}, \bibnamefont{and}
  \bibinfo{author}{\bibfnamefont{D.}~\bibnamefont{Scott}},
  \bibinfo{journal}{Phys. Rev.}
  \textbf{\doi{10.1103/PhysRevD.76.123010}{\bibinfo{volume}{D76}}},
  \doi{10.1103/PhysRevD.76.123010}{\bibinfo{pages}{123010}}
  (\bibinfo{year}{2007}), \eprint{0706.4482} [astro-ph].

\bibitem[{not({\natexlab{a}})}]{note1}
\bibinfo{note}{The details of our treatment of the ISW effect, reionization,
  and polarization do not affect our conclusions, and will be reported
  elsewhere \cite{voidsinprep}.}

\bibitem[{\citenamefont{Lewis et~al.}(2000)\citenamefont{Lewis, Challinor, and
  Lasenby}}]{lcl00notecamb}
\bibinfo{author}{\bibfnamefont{A.}~\bibnamefont{Lewis}},
  \bibinfo{author}{\bibfnamefont{A.}~\bibnamefont{Challinor}},
  \bibnamefont{and} \bibinfo{author}{\bibfnamefont{A.}~\bibnamefont{Lasenby}},
  \bibinfo{journal}{Astrophys. J.}
  \textbf{\doi{10.1086/309179}{\bibinfo{volume}{538}}},
  \doi{10.1086/309179}{\bibinfo{pages}{473}} (\bibinfo{year}{2000}),
  \eprint{astro-ph/9911177};
%
\bibinfo{note}{information on \textsc{camb} is available at
  \href{http://camb.info/}{\tt http://camb.info/}.}

\bibitem[{\citenamefont{Lewis and Bridle}(2002)}]{lb02}
\bibinfo{author}{\bibfnamefont{A.}~\bibnamefont{Lewis}} \bibnamefont{and}
  \bibinfo{author}{\bibfnamefont{S.}~\bibnamefont{Bridle}},
  \bibinfo{journal}{Phys. Rev.}
  \textbf{\doi{10.1103/PhysRevD.66.103511}{\bibinfo{volume}{D66}}},
  \doi{10.1103/PhysRevD.66.103511}{\bibinfo{pages}{103511}}
  (\bibinfo{year}{2002}), \eprint{astro-ph/0205436}.

\bibitem[{not({\natexlab{c}})}]{noteunion}
\bibinfo{note}{Note that the Union dataset assumes a standard $\Lambda$
  background (\eg, in the treatment of outliers), although we do not expect
  this to significantly affect our results.}

\bibitem[{\citenamefont{Reichardt et~al.}(2008)}]{Reichardt:2008ay}
\bibinfo{author}{\bibfnamefont{C.~L.} \bibnamefont{Reichardt}}
  \bibnamefont{et~al.} (\bibinfo{year}{2008}), \eprint{0801.1491} [astro-ph].

\bibitem[{\citenamefont{Jones et~al.}(2006)}]{Jones:2005yb}
\bibinfo{author}{\bibfnamefont{W.~C.} \bibnamefont{Jones}}
  \bibnamefont{et~al.}, \bibinfo{journal}{Astrophys. J.}
  \textbf{\doi{10.1086/505559}{\bibinfo{volume}{647}}},
  \doi{10.1086/505559}{\bibinfo{pages}{823}} (\bibinfo{year}{2006}),
  \eprint{astro-ph/0507494}.

\bibitem[{\citenamefont{Readhead et~al.}(2004)}]{Readhead:2004gy}
\bibinfo{author}{\bibfnamefont{A.~C.~S.} \bibnamefont{Readhead}}
  \bibnamefont{et~al.}, \bibinfo{journal}{Astrophys. J.}
  \textbf{\doi{10.1086/421105}{\bibinfo{volume}{609}}},
  \doi{10.1086/421105}{\bibinfo{pages}{498}} (\bibinfo{year}{2004}),
  \eprint{astro-ph/0402359}.

\bibitem[{\citenamefont{Pryke et~al.}(2008)}]{Pryke:2008xp}
\bibinfo{author}{\bibfnamefont{C.}~\bibnamefont{Pryke}} \bibnamefont{et~al.}
  (\bibinfo{collaboration}{QUaD}) (\bibinfo{year}{2008}), \eprint{0805.1944}
  [astro-ph].

\bibitem[{\citenamefont{Jackson}(2007)}]{Jackson:2007ug}
\bibinfo{author}{\bibfnamefont{N.}~\bibnamefont{Jackson}},
  \bibinfo{journal}{Living Rev. Rel.} \textbf{\bibinfo{volume}{10}},
  \bibinfo{pages}{4} (\bibinfo{year}{2007}), \eprint{0709.3924} [astro-ph].

\bibitem[{\citenamefont{Seo and Eisenstein}(2003)}]{se03}
\bibinfo{author}{\bibfnamefont{H.-J.} \bibnamefont{Seo}} \bibnamefont{and}
  \bibinfo{author}{\bibfnamefont{D.~J.} \bibnamefont{Eisenstein}},
  \bibinfo{journal}{Astrophys. J.}
  \textbf{\doi{10.1086/379122}{\bibinfo{volume}{598}}},
  \doi{10.1086/379122}{\bibinfo{pages}{720}} (\bibinfo{year}{2003}),
  \eprint{astro-ph/0307460}.

\bibitem[{\citenamefont{Eisenstein and Hu}(1998)}]{eh98}
\bibinfo{author}{\bibfnamefont{D.~J.} \bibnamefont{Eisenstein}}
  \bibnamefont{and} \bibinfo{author}{\bibfnamefont{W.}~\bibnamefont{Hu}},
  \bibinfo{journal}{Astrophys. J.}
  \textbf{\doi{10.1086/305424}{\bibinfo{volume}{496}}},
  \doi{10.1086/305424}{\bibinfo{pages}{605}} (\bibinfo{year}{1998}),
  \eprint{astro-ph/9709112}.

\bibitem[{\citenamefont{Clarkson et~al.}(2008)\citenamefont{Clarkson, Bassett,
  and Lu}}]{cbl08}
\bibinfo{author}{\bibfnamefont{C.}~\bibnamefont{Clarkson}},
  \bibinfo{author}{\bibfnamefont{B.}~\bibnamefont{Bassett}}, \bibnamefont{and}
  \bibinfo{author}{\bibfnamefont{T.~H.-C.} \bibnamefont{Lu}},
  \bibinfo{journal}{Phys. Rev. Lett.}
  \textbf{\doi{10.1103/PhysRevLett.101.011301}{\bibinfo{volume}{101}}},
  \doi{10.1103/PhysRevLett.101.011301}{\bibinfo{pages}{011301}}
  (\bibinfo{year}{2008}), \eprint{0712.3457} [astro-ph].

\bibitem[{not({\natexlab{d}})}]{note2}
\bibinfo{note}{Past BAO constraints on void models \cite{gbh08bw08} used an
  isotropized distance measure weighted towards $\Dtbao$, which is related to
  the angular diameter distance, and thus is a weak discriminator of voids when
  fitting to SNe.}

\bibitem[{\citenamefont{Gaztanaga
  et~al.}(2008{\natexlab{a}})\citenamefont{Gaztanaga, Cabre, and Hui}}]{gch08gms08}
\bibinfo{author}{\bibfnamefont{E.}~\bibnamefont{Gaztanaga}},
  \bibinfo{author}{\bibfnamefont{A.}~\bibnamefont{Cabre}}, \bibnamefont{and}
  \bibinfo{author}{\bibfnamefont{L.}~\bibnamefont{Hui}}
  (\bibinfo{year}{2008}{\natexlab{a}}), \eprint{0807.3551} [astro-ph];
%
\bibinfo{author}{\bibfnamefont{E.}~\bibnamefont{Gaztanaga}},
  \bibinfo{author}{\bibfnamefont{R.}~\bibnamefont{Miquel}}, \bibnamefont{and}
  \bibinfo{author}{\bibfnamefont{E.}~\bibnamefont{Sanchez}}
  (\bibinfo{year}{2008}{\natexlab{b}}), \eprint{0808.1921} [astro-ph].

\bibitem[{\citenamefont{Goodman}(1995)}]{goodman95cs07}
\bibinfo{author}{\bibfnamefont{J.}~\bibnamefont{Goodman}},
  \bibinfo{journal}{Phys. Rev.}
  \textbf{\doi{10.1103/PhysRevD.52.1821}{\bibinfo{volume}{D52}}},
  \doi{10.1103/PhysRevD.52.1821}{\bibinfo{pages}{1821}} (\bibinfo{year}{1995}),
  \eprint{astro-ph/9506068};
%
\bibinfo{author}{\bibfnamefont{R.~R.} \bibnamefont{Caldwell}} \bibnamefont{and}
  \bibinfo{author}{\bibfnamefont{A.}~\bibnamefont{Stebbins}},
  \bibinfo{journal}{Phys. Rev. Lett.}
  \textbf{\doi{10.1103/PhysRevLett.100.191302}{\bibinfo{volume}{100}}},
  \doi{10.1103/PhysRevLett.100.191302}{\bibinfo{pages}{191302}}
  (\bibinfo{year}{2008}), \eprint{0711.3459} [astro-ph].

\bibitem[{\citenamefont{Garcia-Bellido and
  Haugboelle}(2008{\natexlab{b}})}]{gbh08b}
\bibinfo{author}{\bibfnamefont{J.}~\bibnamefont{Garcia-Bellido}}
  \bibnamefont{and}
  \bibinfo{author}{\bibfnamefont{T.}~\bibnamefont{Haugboelle}},
  \bibinfo{journal}{JCAP}
  \textbf{\doi{10.1088/1475-7516/2008/09/016}{\bibinfo{volume}{0809}}},
  \doi{10.1088/1475-7516/2008/09/016}{\bibinfo{pages}{016}}
  (\bibinfo{year}{2008}{\natexlab{b}}), \eprint{0807.1326} [astro-ph].

\bibitem[{\citenamefont{Uzan et~al.}(2008)\citenamefont{Uzan, Clarkson, and
  Ellis}}]{uce08}
\bibinfo{author}{\bibfnamefont{J.-P.} \bibnamefont{Uzan}},
  \bibinfo{author}{\bibfnamefont{C.}~\bibnamefont{Clarkson}}, \bibnamefont{and}
  \bibinfo{author}{\bibfnamefont{G.~F.~R.} \bibnamefont{Ellis}},
  \bibinfo{journal}{Phys. Rev. Lett.}
  \textbf{\doi{10.1103/PhysRevLett.100.191303}{\bibinfo{volume}{100}}},
  \doi{10.1103/PhysRevLett.100.191303}{\bibinfo{pages}{191303}}
  (\bibinfo{year}{2008}), \eprint{0801.0068} [astro-ph].

\bibitem[{\citenamefont{Benitez et~al.}(2008)}]{pau08}
\bibinfo{author}{\bibfnamefont{N.}~\bibnamefont{Benitez}} \bibnamefont{et~al.}
  (\bibinfo{year}{2008}), \eprint{0807.0535} [astro-ph].

\bibitem[{not({\natexlab{e}})}]{noteboss}
\bibinfo{note}{\href{http://cosmology.lbl.gov/BOSS/}{\tt
  http://cosmology.lbl.gov/BOSS/}}.

\end{thebibliography}

\end{document}